\documentclass[baaa]{baaa}

\usepackage[pdftex]{hyperref}
\usepackage{subfigure}
\usepackage{natbib}
\usepackage{helvet,soul}
\usepackage[font=small]{caption}

\contriblanguage{1}

\contribtype{1}

\thematicarea{1}

\title{Redshift Horizon for the {\LARGE\slshape{Origins Space Telescope}}}
\subtitle{from Primordial Dust Emission}

\titlerunning{Redshift Horizon for the {\large\slshape{Origins Space Telescope}}}

\author{M.E. De Rossi\inst{1,2} \& V. Bromm\inst{3}}
\authorrunning{De Rossi \& Bromm}

\contact{mariaemilia.dr@gmail.com}

\institute{
Universidad de Buenos Aires, Facultad de Ciencias Exactas y Naturales y Ciclo B\'asico Com\'un. Buenos Aires, Argentina
\and CONICET-Universidad de Buenos Aires, Instituto de Astronom\'{\i}a y F\'{\i}sica del Espacio (IAFE). Buenos Aires, Argentina
\and Department of Astronomy, University of Texas at Austin, 2511 Speedway, Austin, TX 78712, USA
}

\resumen{
	Exploramos la posibilidad de detectar las primeras galaxias con el telescopio
	de próxima generación {\sl Origins Space Telescope} (OST) aplicando un modelo analítico
	de emisión de polvo primordial. Mediante el análisis de las densidades de fuentes
	como función del corrimiento al rojo ($z$), y considerando exposiciones de campo profundo
	con el {\sl Origins Survey Spectrometer}, estimamos que el
	horizonte de corrimiento al rojo para detectar una fuente individual estaría por arriba 
	de $z\sim7$ para sistemas con cocientes polvo-metal mayores que los esperados para galaxias 
	primigenias típicas. 
	Por otro lado, si los límites de confusión pudieran ser afrontados, el instrumento 
	{\sl Far-infrared Imager and Polarimeter} permitiría la detección de sistemas más débiles y típicos a $z>7$. 
Dada la dependencia de los resultados obtenidos con las propiedades del polvo primigenio,
concluimos que el OST podría proveer claves importantes sobre la naturaleza del medio interestelar
en el Universo temprano.
	}

\abstract{
We explore the possibility of detecting the first galaxies with the next generation
{\sl Origins Space Telescope} (OST) by applying an analytical model of primordial
dust emission. By analysing source densities as a function of redshift ($z$), and considering
deep-field exposures with the {\sl Origins Survey Spectrometer},
we estimate that the redshift horizon for detecting one
individual source would be above $z\sim 7$ for systems with dust-to-metal ratios
higher than those expected for typical
primeval galaxies. On the other hand, if confusion limits could be overcome, the
{\sl Far-infrared Imager and Polarimeter} would enable the detection
of more typical fainter systems at $z>7$. 
Given the dependence of the obtained results with the properties of primeval dust,
we conclude that the OST could provide important clues to the nature of the interestellar
medium in the early Universe.
}

\keywords{galaxies: high-redshift --- galaxies: evolution ---
galaxies: formation --- galaxies: star formation ---
cosmology: theory
}

\begin{document}

\maketitle

\section{Introduction}
\label{sec:Introduction}
Primeval galaxies at $z\gtrsim7$ are promising targets for current and upcoming
observational facilities operating at different wavelengths, such as the {\sl James Webb Space
Telescope} (JWST), the Atacama Large
Millimeter/sub-millimeter Array (ALMA) and the Square Kilometre Array
(SKA), among others.
In this context, future space-borne FIR telescopes
could play a crucial role on exploring the nature and properties of the first
dust-emitting galaxies at the very dawn of star formation \citep[e.g.][]{derossi2019}.
These systems ($z\gtrsim7$) are expected to have very low dust densities
and FIR fluxes, being their detection below the capabilities of
current and near future observatories \citep[e.g.][]{derossi2017}.

\begin{figure*}[!h]
  \centering
  \includegraphics[width=0.40\textwidth]{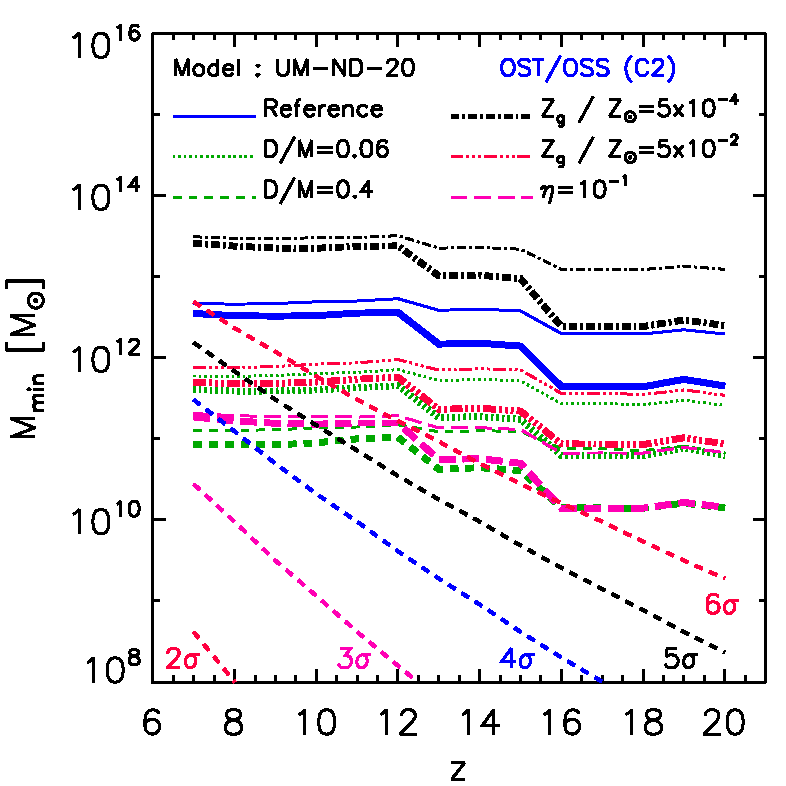}
  \includegraphics[width=0.40\textwidth]{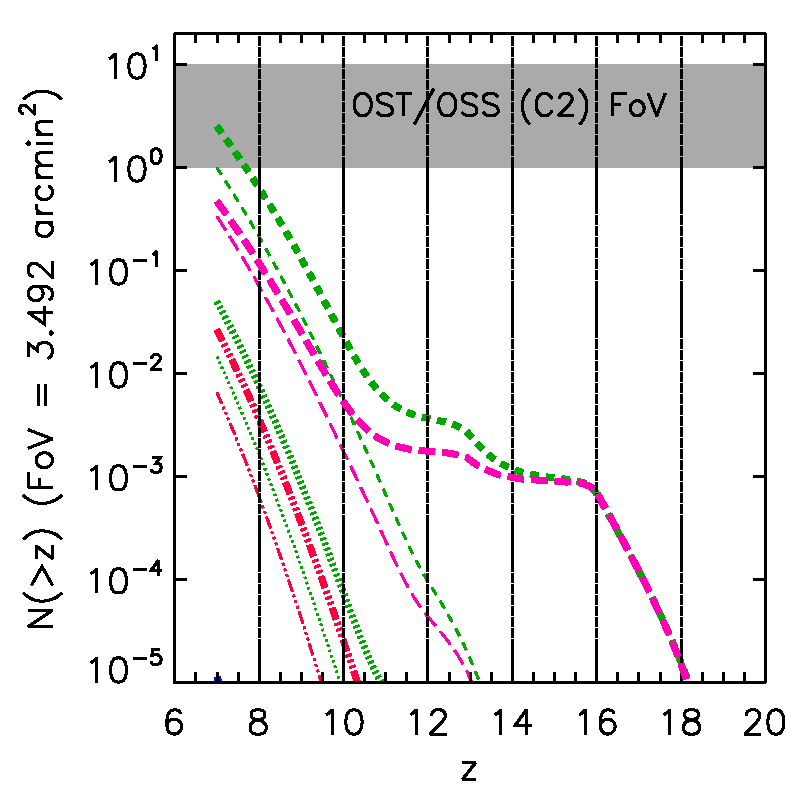}
  \caption{
	  {Impact of variations of model parameters on detections with OSS (band 6). 
	  Results correspond to a deep-field
	exposure ($10^6~{\rm s}$) within a FoV.
Left panel: $M_{\rm min}-z$ relation. 
For comparison, different $\nu$-$\sigma$ peaks corresponding to the adopted
cosmological model are plotted.
Right panel: $N-z$ relation.
In both panels, thin and thick lines are associated with the standard and shock size distributions, respectively.
The reference model (solid blue line) corresponds to our standard parameters.
The effects of variations of model parameters with respect to the standard model are shown with different line styles.
In the right panel, the curves that are not shown lie below the plotted region.
	}}
  \label{fig:OSS}
\end{figure*}

\citet[][]{derossi2019} explore in detail the possibility of detecting first
galaxies during FIR/sub-mm surveys with a generic FIR telescope. 
In this work, we implement their methodology to evaluate the possibility of 
detecting primeval FIR/sub-mm sources at 
$z \gtrsim 7$ with the future {\sl Origins Space Telescope} (OST), planned for launch in the 
2030s.
The OST further evolves the concept mission study for
the {\em Far-Infrared Surveyor}.
Among the key science goals for this facility will be the cosmic origin of dust, 
elucidating the first sources of dust emission, which are beyond the current horizon of observability, 
given the sensitivity of existing instruments.
The OST is expected to attain sensitivities 100-1000 times greater 
than any previous FIR mission
achieving unprecedented depths to study the most distant galaxies.
In this manuscript, we focus on the detectability of FIR/sub-mm dust continuum emission
from primeval galaxies with the following instruments: the {\sl Origins Survey Spectrometer} (OSS, concept 2-C2)
and the {\sl Far-infrared Imager and Polarimeter} (FIP, C2).
For more details and related documetation regarding the OST, see https://asd.gsfc.nasa.gov/firs/ and
https://origins.ipac.caltech.edu/.

\begin{figure*}[!h]
  \centering
  \includegraphics[width=0.40\textwidth]{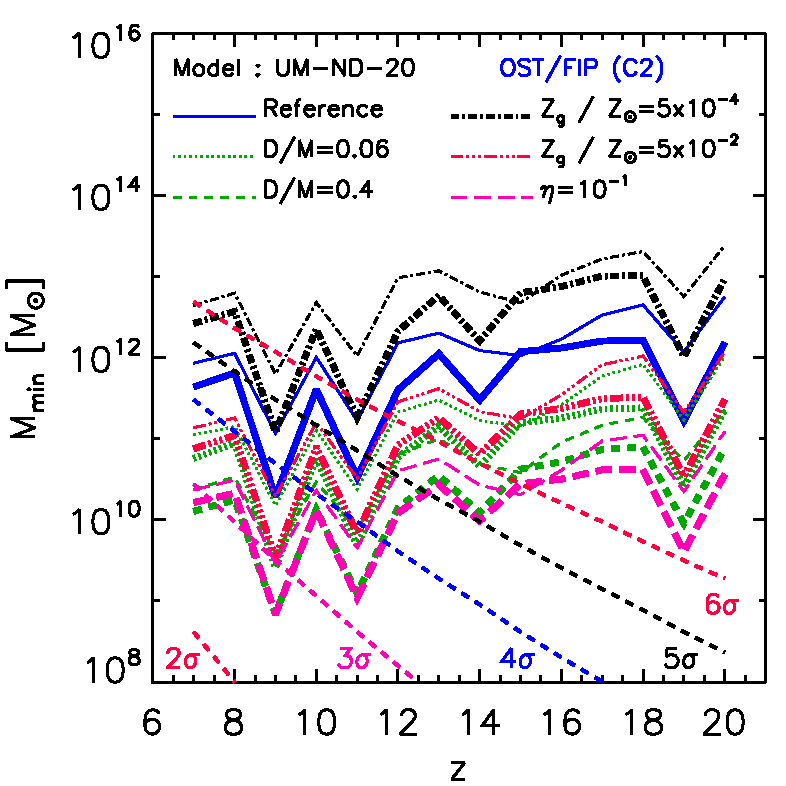}
  \includegraphics[width=0.40\textwidth]{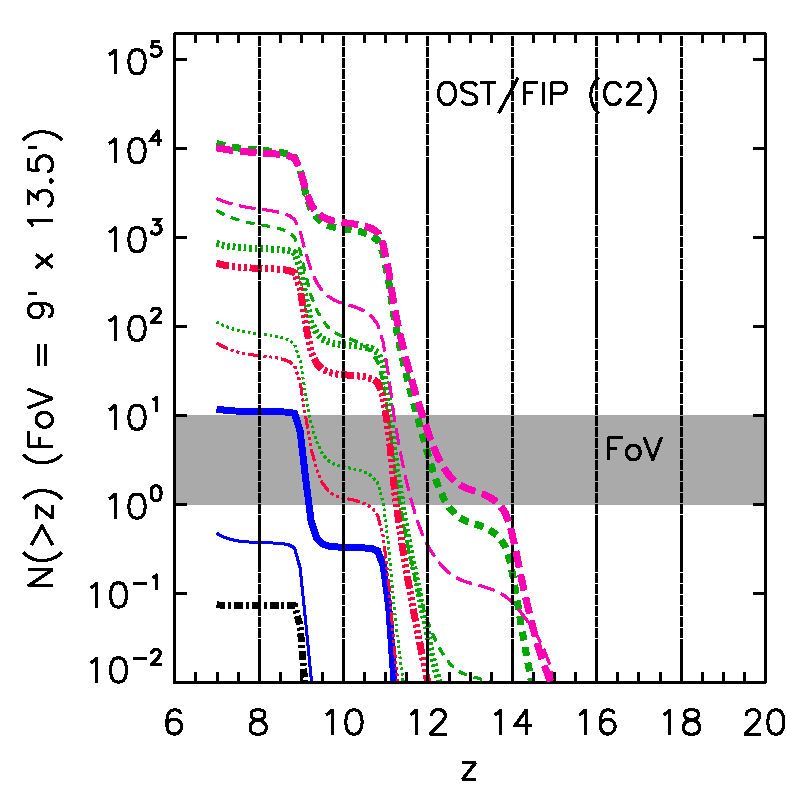}
  \caption{
	  Similar to Fig.~\ref{fig:OSS}, but for FIP at $250~{\mu}{\rm m}$ and during a deep-field exposure
	    within a FoV.
}
  \label{fig:FIP}
\end{figure*}

\section{Dust model}
Dust emission was estimated using the methodology described in \citet{derossi2017} and \citet{derossi2019},
which has proven to be useful for studying high-$z$ galaxies \citep{derossi2018}.
We referred the reader to those papers for a detail description
of the dust model;  here, we only present a brief summary of it.

A model galaxy consists of a dark matter halo hosting a central cluster of Pop~II stars, surrounded by a mixed
phase of gas and dust. 
Our standard model assumes a dust-to-metal mass ratio $D/M = 5 \times 10^{-3}$, a gas metallicity of
$Z_{\rm g}=5\times10^{-3}~{\rm Z}_{\odot}$ and a star formation efficiency of $\eta=0.01$, which are typical values for
first galaxies (\citealt{greif2006, mitchellwynne2015, schneider2016}; see
\citealt{derossi2017, derossi2019} for more details regarding model parameters and the effects of their variations). The spectral energy distribution associated to stars was obtained from {\sc YGGDRASIL} model grids
\citep[][]{zackrisson2011}.
We considered different silicon-based dust models given in \citep[][]{cherchneff2010}. 
However, for the sake of clarity, we only present results corresponding to the
so-called UM-ND-20 model (see \citealt{derossi2017}, for details).
We have checked that other chemical compositions of dust lead to similar general trends.
For the grain-size distribution, we adopted the
‘standard’ and ‘shock’ prescriptions used in \citet{ji2014}.
Dust temperature ($T_{\rm d}$) was determined assuming thermal equilibrium and dust emissivity
was estimated by applying the Kirchhoff’s law for the estimated $T_{\rm d}$ profile.

By comparing the sensitivity of the OST instrument with the
average observed fluxes of galaxies at a given wavelength band, we determine
the lowest virial mass ($M_{\rm min}$) that a galaxy should have to be detected at a given $z$.
The redshift horizon ($z_{\rm lim}$)
is defined as the highest $z$ above which the projected number of detected sources
per given solid angle ($\Delta \Omega$)
is $N \le N_{\rm crit}$, where we here
consider $N_{\rm crit} = 1-10$ \citep[see][]{derossi2019}. 

\section{Results}
For OSS and FIP, the
number of detected sources, $N$, was obtained assuming deep-field exposures ($10^6$\,s) 
within each instrument field-of-view (FoV).
Deep-field sensitivities were obtained by scaling the expected sensitivities for 1\,hr
with the square root of the integration time ($10^6$~s).
Details regarding the FoV and sensitivity (1\,hr) of each instrument were 
obtained from \cite{meixner2019}.

In the case of OSS, we estimated $M_{\rm min}$ and $N$ for a deep field
exposure in its band 6 ($\lambda = 336 - 589~\mu{\rm m}$). We adopt
a sensitivity of $2.34~\mu {\rm Jy}$, corresponding to an integration time
of $10^6~{\rm s}$, and a FoV of $3.492~{\rm arcmin}^2$ \citep{meixner2019}.
As can be seen in Fig.~\ref{fig:OSS}, 
in the case of a very low metallicity of $Z_{\rm g}=5\times10^{-4}~{\rm Z}_{\sun}$,
$M_{\rm min}$ ranges between
$\sim 10^{12}~{\rm M}_{\sun}$ ($z\approx20$) and $\sim 10^{14}~{\rm M}_{\sun}$ ($z\approx7$),
with the exact values depending on the size distribution of dust grains.
For high dust-to-metal ratios ($D/M=0.4$) or star formation
efficiencies ($\eta = 0.1$), $M_{\rm min}$ ranges between
$\sim 10^{10}~{\rm M}_{\sun}$ ($z\approx20$) and $\sim 10^{11}~{\rm M}_{\sun}$ ($z\approx7$),
also, with the exact values depending on the size distribution of dust grains.
All other assumed parameters predict an intermediate behaviour.
Note that, for OSS, $M_{\rm min}$ decreases towards its minimum value at $z\gtrsim 15$.
This is caused by the very strong negative K-correction that affects first galaxies at FIR/sub-mm 
wavelengths, as discussed in detail in \citet{derossi2019}
We also see that OSS would reach the redshift horizon at $z\approx7-8$ only in the case of 
very high dust-to-metal ratios.

In the case of FIP (unlike the case of OSS), the characteristic wavelength 
($\approx 250~{\mu}{\rm m}$) is below the peak of
dust emission of our high-$z$ galaxies ($\approx 500~{\mu}{\rm m}$, \citealt{derossi2017,derossi2019} ), so their detection is not favoured
by the strong negative K-correction that we mentioned before.
In addition, FIP will likely be affected by confusion noise during deep 
exposures. If confusion limits could be overcome
(e.g. by applying multi-wavelength analysis), we show below that its high sensitivity would 
enable the detection of dust continuum emission from {\em typical} first galaxies.
For FIP, we adopt a sensitivity of $0.15~{\mu}{\rm Jy}$ at $250~\mu{\rm m}$, corresponding to an
exposure of $10^6~{\rm s}$, and a FoV of $9' \times13.5'$ \citep{meixner2019}. As shown in Fig.~\ref{fig:FIP}, for a very 
low metallicity of $Z_{\rm g}=5\times10^{-4}~Z_{\sun}$, $M_{\rm min}$ ranges between
$\sim 10^{13}~{\rm M}_{\sun}$ ($z\approx20$) and $\sim 10^{12}~{\rm M}_{\sun}$ ($z\approx7$), 
with the exact value depending on the grain size distribution. In the case of high dust-to-metal 
ratios ($D/M = 0.4$) or star formation
efficiencies ($\eta = 0.1$), $M_{\rm min}$ ranges between
$\sim 10^{11}~{\rm M}_{\sun}$ ($z\approx20$) and $\sim 10^{9}~{\rm M}_{\sun}$ ($z\approx7$). Other parameter choices predict an intermediate behavior.
With respect to the redshift horizon,
our reference model predicts $N = 1-10$ at $z\approx7$ and $z\approx9$, for the
standard and shock size distribution, respectively. In the case
of $Z_{\rm g}=5\times10^{-4}~Z_{\sun}$,
$N(>7)<1$; thus, a wider area of the sky should be searched to increase the probability
of detecting low metallicity sources. For higher $Z_{\rm g}$ and $D/M$, the horizon moves towards 
higher redshifts. For our extreme case of $\eta \approx 0.1$ and a shock grain size distribution, 
the horizon is reached at $z=12-14$, with other parameter values leading to intermediate behavior.

\section{Conclusions}
We explore the prospects of detecting the dust
continuum emission from first galaxy populations with the next-generation {\sl Origins Space Telescope}.
We focus on OSS and FIP (concept 2) instruments.
We consider deep-field exposures adopting
similar integration times ($10^6~{\rm s}$) within each instrument field-of-view (FoV).

During deep-field exposures within a FoV, OSS may be able to detect
the dust continuum emission of massive sources with very high dust fractions
($D/M \gtrsim 0.4$).
The more elusive fainter galaxies could be observed by gravitational lensing effects.
In the case of FIP, deep FoV exposures could allow the detection of 
more {\em typical}
primeval galaxies with low metallicities ($\sim 0.005~Z_{\sun}$) and
dust-to-metal ratios ($\sim 0.005$). However, such efforts would require the development of 
techniques that help to overcome
confusion noise and remove interlopers at $z<7$.  

Taking into account the sensitivity of the OST redshift horizon to the assumed
dust properties, measurements of FIR/sub-mm source densities at $z>7$
could provide important clues to the nature and origin of the first sources of
primordial dust in the early Universe.

\begin{acknowledgement}
MEDR thanks the Asociación Argentina de Astronomía for providing
with partial financial support for attending its 61st annual meeting.
MEDR is grateful to PICT-2015-3125 of ANPCyT (Argentina) and also to
Mar\'{\i}a Sanz and Guadalupe Lucia for their help and support.
VB acknowledges support from NSF grant AST-1413501.
We thank Alexander Ji for providing tabulated dust opacities for the different
dust models used here.
This work makes use of the Yggdrasil code \citep{zackrisson2011}, which adopts
Starburst99 SSP models, based on Padova-AGB tracks \citep{leitherer1999, vazquez2005}
for Population~II stars.
\end{acknowledgement}


\bibliographystyle{baaa}
\small
\bibliography{bibliografia}
 
\end{document}